\documentclass[
  journal=pasa,
  manuscript=research-paper, 
  year=202X,
  volume=YY,
]{cup-journal}

\usepackage{hyperref}
\usepackage{amsmath}
\usepackage{multirow} 
\usepackage{amssymb,microtype,siunitx,booktabs}
\title{Positive AGN feedback in the outskirts of nearby barred spiral galaxies?}

\author{Bannanje Ananthamoorthy}  
\affiliation{Manipal Centre for Natural Sciences, Manipal Academy of Higher Education, Karnataka, Manipal, 576 104, India}

\author{Debbijoy Bhattacharya} 
\affiliation{Manipal Centre for Natural Sciences, Manipal Academy of Higher Education, Karnataka, Manipal, 576 104, India}
\email[D. Bhattacharya]{debbijoy.b@manipal.edu}

\author{P. Sreekumar}  
\affiliation{Manipal Centre for Natural Sciences, Manipal Academy of Higher Education, Karnataka, Manipal, 576 104, India}

\received {dd Mmm YYYY}
\revised  {dd Mmm YYYY}
\accepted {dd Mmm YYYY}
\published{22 September 202X}

\begin{document}

\begin{abstract}

Observational evidence regarding the impact of AGN feedback on star formation (SF) in non-jetted galaxies is limited. With the available high-resolution UV observations from {\sl AstroSat}-UVIT, complemented by {\sl GALEX}, we studied the SF properties in the outskirts ($>0.5R_{25}$) of six AGN-host galaxies and compared them with four non-AGN galaxies of similar morphology. We observed a higher SF rate density ($\Sigma_{\text{SFR}}$) for the UV knots in AGN-host galaxies, and it falls off less rapidly compared to non-AGN galaxies, suggesting positive AGN feedback in the outskirts of AGN-host galaxies. Additionally, FUV attenuation (A$_{\text{FUV}}$) is also enhanced in the outer regions and falls less rapidly in AGN-host compared to non-AGN, indicating that the feedback could be coupled with dust. We speculate that the radiation-pressure-driven and/or wind mode AGN feedback could be at play even in low-luminosity nearby AGN-host galaxies.

\end{abstract}

\section{Introduction} \label{sec:intro}
The energy output from the Active Galactic Nucleus (AGN) can influence the evolution of the host galaxy. A small fraction of energy ($\sim1\%$) from AGN, if coupled with the surroundings, can exceed the gas binding energy in the galaxies \citep[e.g., ][]{Fabian2012, Silk1998, Ostriker2005}. AGN feedback models are often invoked to explain the steep decline in the galaxy mass function \citep[e.g.,][]{Bower2006, Croton2006}, star formation (SF) quenching in massive galaxies \citep[e.g., ][]{Schawinski2007, Springel2005}, and preventing SF in cooling flows of galaxy clusters \citep[see e.g.,][for review]{McNamara2007}. Furthermore, the correlation between the black hole mass (M$_{BH}$) and the galaxy bulge velocity dispersion \citep[$\sigma_b$;][]{Gebhardt2000, Ferrarese2000}, and similar evolutionary history of black hole growth and SF \citep[e.g.,][and references therein]{Kormendy2013} suggests that the BH and the SF in a galaxy may co-evolve.

The energy output from the AGN can interact with a medium via winds, jets, or radiation pressure on dust/gas embedded in the galaxy \citep[e.g.,][and references therein]{Fabian2012}. The impact of radiative mode AGN feedback on SF is not well understood, neither theoretically nor observationally. It is due to a poor understanding of the physical processes responsible for the coupling of AGN radiation with the surrounding medium \citep{Costa2018, King2015}. AGN radiation can interact with the environment via winds, Compton heating, or interaction with dust \citep{Costa2018, King2015, Ishibashi2012}.

Simulations suggest that wind mode AGN feedback can suppress SF in nuclear regions (within 100 pc) by a factor of 10, and on a large scale (a few kpc), the outflow materials may not escape the host galaxy \citep{Hopkins2016}. Simulations investigating the effect of Compton heating \citep{Kim2011} suggest that the central region ($\sim$ within 0.1-0.2 kpc) will be devoid of molecular gas, as the medium is hot and turbulent, resulting in suppressed SF. However, the outskirts of the galaxy remain unaffected. Simulations/semi-analytical models of radiation pressure-driven AGN feedback coupled with dust suggest that it can drive shells of gas to the outskirts of the galaxy, and stars can form in outflowing shells obscured by dust \citep{Ishibashi2012, Ishibashi2013, Costa2018}. \citet{Ishibashi2013} also suggested that a significant amount of dusty gas can spread over the entire galaxy. Simulations by \citep{Costa2018} indicate that the radiation pressure-driven dusty outflows will not escape the galaxy even for a high-luminous AGN.

Radiation pressure-driven AGN feedback is considered as the possible reason for inside-out growth, typically observed in early-type galaxies at redshift $\sim 2$ \citep{VanDokkum2010}. The outflows observed in several galaxies \citep[e.g.,][]{Carniani2015} are believed to be due to the radiative mode AGN feedback. However, evidence of its impact on the SF is limited. Studies on a sample of galaxies yields contrasting results regarding the AGN feedback effect on the global star formation rate (SFR) \citep[e.g.,][and references therein]{Guo2019, Harrison2024}. These contrasting results could be due to the relative displacement of the SF from the inner to the outer region without much change in the overall SF, as suggested in simulations \citep[e.g.,][]{Robinchaud2017}. Therefore, studying resolved SF in the galaxies is essential in understanding the effect of AGN feedback. Most of the studies related to spatially resolved SF focus on the inner few kpc scales \citep{Bing2019, George2018, Joseph2022}, and studies extending to outer regions of the galaxy are very limited \citep[e.g.,][]{Guo2019, Acharya2024, Sanchez2018}. Based on H$\alpha$ emission, \citet{Guo2019} suggested that AGN-host galaxies have a lower SFR surface density ($\Sigma_{\text{SFR}}$) at all radial bins compared to non-AGN. \citet{Acharya2024} and \citet{Sanchez2018} observed a similar radial profile for $\Sigma_{\text{SFR}}$ in AGN and non-AGN galaxies. The $\Sigma_{\text{SFR}}$ is best derived using the individual knot areas obtained from high angular resolution observations. However, previous studies are limited by angular resolution, which restricts them to probe scales of 1 kpc or higher. The study of SF in AGN and non-AGN galaxies, probing the scale of individual stellar clusters, is still missing. Such a spatially resolved study would allow us to probe SF in galaxies unaffected by large-scale morphological features, such as spiral arms in the galaxy. Recently, \citet{Nandi2024} carried out a study of seven Seyfert galaxies using the UltraViolet Imaging Telescope (UVIT) onboard {\sl AstroSat}, and observed that the nuclear to total SFR ratio of these galaxies correlates with the Eddington ratio of the galaxies. \citet{Ananthamoorthy2025} studied one of the nearby galaxies, NGC 1097, and suggested that the $\Sigma_{\text{SFR}}$ in the nuclear region of the galaxy might have been enhanced due to the AGN positive feedback. However, in these works, the large-scale effect of AGN on SF was not probed.

\begin{figure}
\centering
	\includegraphics[width=8cm]{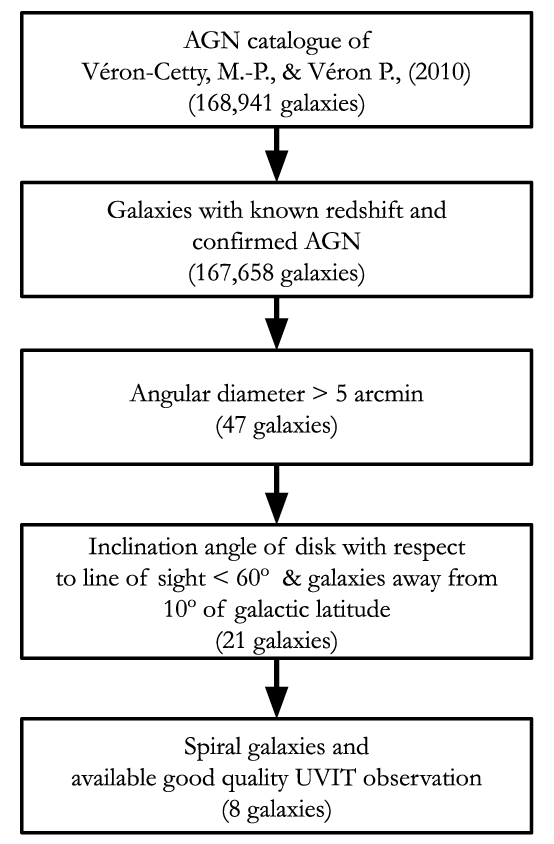}
    \caption{The flowchart for the criteria used for selection of AGN sample.}
    \label{fig:Sample_selection}
\end{figure}

In this work, we used high-resolution ($1.2-1.7^{\prime\prime}$) UV observations from the UVIT onboard {\sl AstroSat} \citep{Kumar2012, Tandon2017, Tandon2020} to study spatially resolved SF (scale of $\sim 50-100$ pc) of nearby barred AGN-host galaxies and compared it with a sample of non-AGN galaxies. We restricted the investigation to the outskirts \citep[$>0.5$R$_{25}$; R$_{25}$ is the isophotal radius corresponding to $25$ mag/arcsec$^{2}$ in B band;][]{RC3} of the galaxy as other galaxy components, such as the bar, can drive the SF in the galaxy's inner regions, which can vary significantly between galaxies. Moreover, theoretical predictions of the effect on SF by different modes of radiative AGN feedback are differentiable in the outer regions of the galaxies \citep{Hopkins2016, Kim2011, Ishibashi2013}. Therefore, the study of SF properties on the outskirts of the galaxy can provide crucial insights into the operational mode of feedback in nearby galaxies.

\section{Sample and Data}

\subsection{Sample}
Our AGN sample is derived from the AGN catalogue of \citet{Veron2010}. Galaxies with known redshift and confirmed AGN are considered. We obtained physical properties of the galaxies, such as angular size, inclination angle (angle of line-of-sight with respect to the disk normal), morphology, etc., from the HyperLeda\footnote{http://leda.univ-lyon1.fr/} database \citep{Makarov2014}, NASA/IPAC Extragalactic Database (NED)\footnote{https://ned.ipac.caltech.edu/} or SIMBAD \footnote{https://simbad.cds.unistra.fr/simbad/} database. 
\begin{figure}
\centering
	\includegraphics[width=9.5cm]{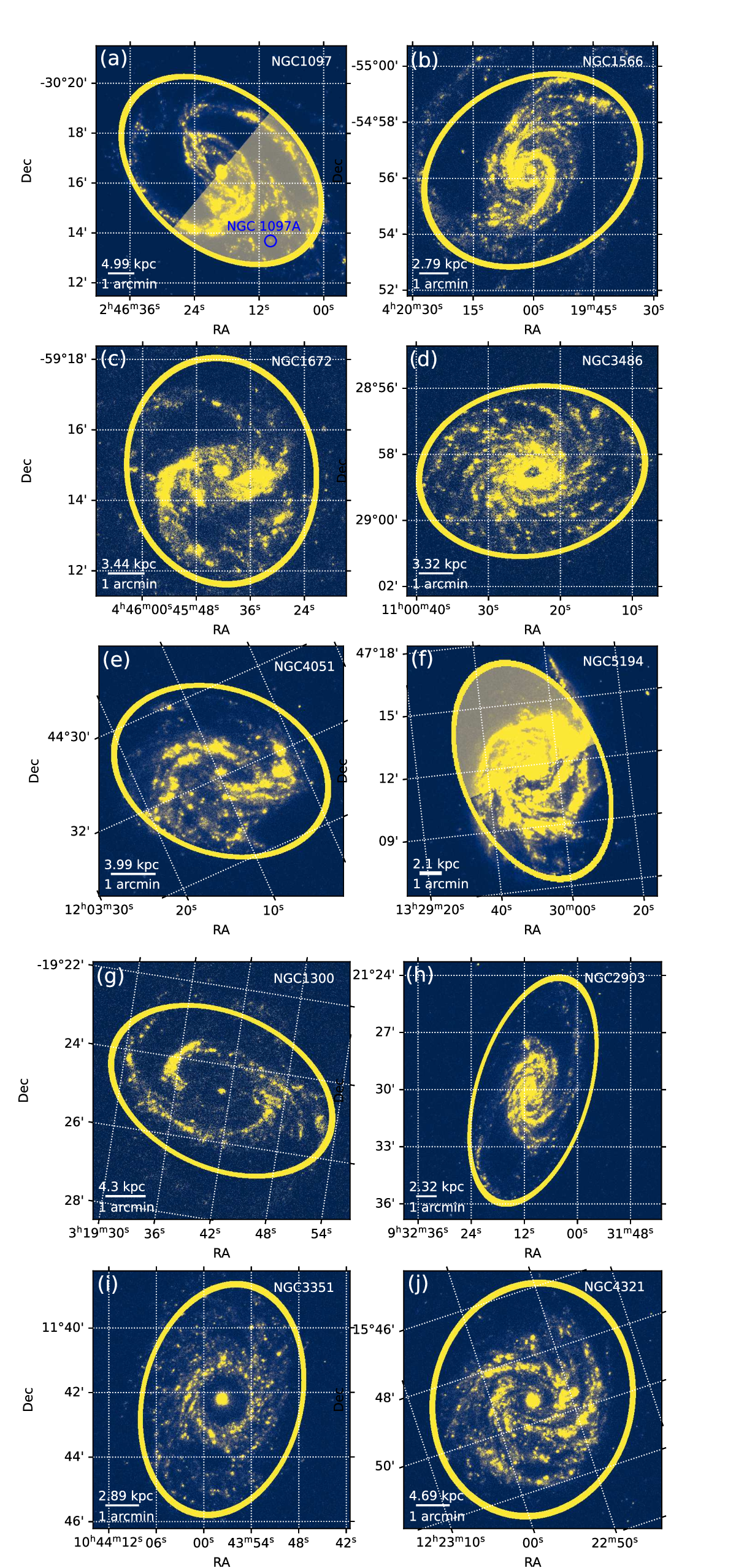}
    \caption{UVIT images of galaxies in the sample. The first three rows correspond to AGN-host galaxies, and the last two rows correspond to non-AGN galaxies. The ellipse in each image represents the galaxy's optical radius (R$_{25}$). For NGC 1097 (Panel (a)), the position of the companion galaxy, NGC 1097A, is marked in the blue circle. For NGC 1097 and NGC 5194 (Panels (a) and (f)), the shaded elliptical region corresponds to the region affected by the merger. These masked regions are excluded from the study of AGN's impact on these galaxies (see Section-\ref{sec:Analysis}).}
    \label{fig:UVIT_obs}
\end{figure}
We considered galaxies with an angular diameter of $> 5^\prime$ to facilitate a spatially resolved study of the SF properties. We selected only galaxies with inclination angle $< 60^{\circ}$ and located away from the Galactic plane ($|b|>10^{\circ}$). The cutoff on inclination angle was applied to minimise the effect of attenuation of the UV emission intrinsic to the host galaxy, and the Galactic latitude limit was imposed to avoid severe extinction from our Galaxy. As elliptical galaxies have little ongoing SF and are typically UV faint, we considered only spiral galaxies for the current study. Also, the galaxies with good quality FUV observations from {\sl AstroSat}-UVIT are considered for the study. The flowchart of the criteria used for AGN sample selection with the number of galaxies present at each step is provided in Figure~\ref{fig:Sample_selection}. We noticed that two galaxies, NGC 4450 and NGC 4579, did not have any detected UV knots beyond 0.7 R$_{25}$. As this study focuses mainly on the possible effect of AGN feedback at the outskirts of the galaxy, we excluded these galaxies from further analysis. Finally, we have six AGN-host galaxies in our sample. All the galaxies are barred and of morphological type between b and c types. As the bars can significantly impact the SF of the galaxy, a careful selection of control galaxies (also having bars) is essential. For the control sample, we searched for the galaxies in the HyperLeda, NED and SIMBAD databases (a) classified as non-AGN based on earlier published results, (b) with angular size $> 5^{\prime}$, (c) inclination angle $< 60^{\circ}$, (d) having a bar, (e) morphological class similar to those of AGN galaxies, and (f) availability of UVIT observation. We could obtain four non-AGN galaxies as a control sample.

In \citet{Nandi2024}, NGC 4321 was considered as an AGN-host/transition galaxy. However, using X-ray and other multiband data, \citet{Martin2009} suggested that this galaxy does not show any evidence for the presence of AGN. Therefore, we have considered this galaxy in a control sample for this study. The physical properties of the AGN host and control galaxies in the sample are provided in Table~\ref{tab:gal_prop}.

\begin{table*}
	\centering
	\caption{Details of the galaxy properties in our sample. \label{tab:gal_prop}}
     
     \footnotesize
	\begin{tabular}{cccccccccccccc} 

		\hline
		Name & R.A. & Dec. & Morph. & R$_{25}$  &Distance  & P.A.  & e   & $\log\left(\frac{M_{BH}}{M_\odot}\right)$ & $\log(\text{L}_{bol})$ & AGN &$M_{BH}$/Lum\\
           (NGC)  & deg. & deg. & & $^{\prime \prime}$ (kpc) & Mpc  & deg. & b/a   & &AGN & Type &Ref. \\
           (1) & (2) & (3) & (4) & (5) & (6) & (7) & (8) & (9) & (10) & (11) & (12)  \\
		\hline
            \multicolumn{11}{l}{AGN-host galaxies} \\
            \hline
            1097 & $41.579410$  & $-30.27491$   &SBb    & $280~(23.3)$     & $17.16$   & $130$     & $0.68$    &$8.15^{+0.08}_{-0.11}$  &$42.279$       &S3b    & a/k \\
            \hline
            1566 & $65.001642$  & $-54.937944$  &SABb   & $250~(11.6)$   & $9.587$   & $60$      & $0.79$    &$6.92^{+0.15}_{-0.22}$  &$42.717$       &S1.5   & b,c/k \\
            \hline
            1672 & $71.427093$  & $-59.247267$  & SBb   & $198~(11.3)$   & $11.809$  & $171$     & $0.83$    &$7.7\pm0.1$              & $40.412$      &S      & d/l,k \\
            \hline
            3486 & $165.09956$  & $28.975114$   & SBc   &$212~(11.7)$      & $11.398$  & $80$      & $0.74$    & $6.14^{+0.10}_{-0.13}$  &$40.37$        & S2    & e,c/m,k\\
            \hline
            4051 & $180.790030$ & $44.531308$   & SABb  &$157~(10.4)$   & $13.714$  & $135$     & $0.74$    & $6.24^{+0.17}_{-0.15}$ &$42.786$       & S1n   &f/k\\
            \hline
            5194 & $202.469626$ & $47.195207$   & SABb  &$337~(11.8)$    & $7.225$   & $163$     & $0.62$    & $6.95^{+0.18}_{-0.31}$ &$41.741$       &S2     & b,c/k \\
            \hline
            \multicolumn{11}{l}{Non-AGN Galaxies}\\
            \hline
            1300 & $49.920939$  & $-19.411153$  &SBbc   &$185~(13.3)$      &$14.789$   &$ 106$     & $0.66$    &  $7.82\pm0.29$          &...            & ...   & g,h\\
            \hline
            2903 & $143.04213$  & $21.50083$    &SBbc   & $378~(14.6)$   & $7.990$   & $17$      & $0.48$    & $6.96^{+0.08}_{-0.09}$  &...            & ...   & i,h\\
            \hline
            3351 & $160.990603$ & $11.703740$   & SBb   &$222~(10.7)$    & $9.940$   & $13$      & $0.68$    & $<6.96$                  &...            &...    & j\\
            \hline
            4321 & $185.72859$  & $15.821635$   & SABb  &$222~(17.4)$    & $16.117$  & $30$      & $0.85$     &$6.47^{+0.08}_{-0.09}$   &...            &...    & i,h\\
            
             \hline
	\end{tabular}

    \vspace{0.1ex}

 {\raggedright \textsc{Notes.--} Col.(1): Name of the galaxy. Col.(2) and Col.(3): Right ascension and Declination (J2000; from NED). Col.(4): Morphology (from HyperLeda). Col.(5): Optical Radius \citep[][NED]{RC3} in arcsec and kpc. Col.(6): Distance (from NED). Col.(7): Position Angle of the galaxy measured from the north \citep[][NED]{RC3}. Col.(8): Ratio of minor to major axis \citep[][NED]{RC3}. Col.(9): BH mass. For NGC 3486, we used the M$_{BH}$ vs. $\sigma_b$ relation by \citet{Woo2002} and the velocity dispersion values from \citet{Barth2002} to derive M$_{BH}$.  Col.(10): Log of AGN bolometric luminosity (in ergs s$^{-1}$). For NGC 3486 and NGC 1672, we used the relation by \citet{Spinoglio2024} to convert the 2-10 keV X-ray luminosity to AGN bolometric luminosity (L$_{bol,AGN}$). Col.(11): AGN Type \citep{Veron2010}. S: Unclassified Seyfert; S1.5: Seyfert-1.5; S1n: narrow-line Seyfert-1 galaxy. S2: Seyfert-2; S3b: Low Ionisation Nuclear Emission Region (LINER) with broad Balmer lines. Col.(12) Reference for BH mass and AGN luminosity. \\References: a: \citet{Onishi2015}; b: \citet{Nelson1995}; c: \citet{Woo2002}; d: \citet{Combes2019}; e: \cite{Barth2002}; f: \citet{Denney2010}; g: \citet{Atkinson2005}; h: \citet{Berrier2013}; i: \citet{Ferrarese2002}; j: \citet{Sarzi2002}; k: \citet{Spinoglio2024}; l: \citet{Jenkins2011}; m: \citet{Annuar2020}. \par}

\end{table*}

\begin{table*}
	\centering
	\caption{Details of {\sl AstroSat}-UVIT and {\sl GALEX} observations \label{tab:obs}}

\begin{tabular}{lcccc cccc} 
\hline
Name & \multicolumn{4}{c}{{\sl AstroSat}-UVIT} & & \multicolumn{3}{c}{{\sl GALEX}} \\
\cmidrule(){2-5}
\cmidrule(){7-9}

		& Observation ID  & Filter & Exposure &FWHM$^{\text{a}}$&& Tile name & FUV Exp. & NUV Exp. \\
         (NGC)    & & & Time (sec.) & arcsec & & &(sec.)&(sec.)   \\
		\hline
        \multicolumn{8}{l}{AGN-host galaxies} \\
        \hline
            1097& A10\_116T01\_9000004082  & F154W &$30881.8$& $1.2$& &NGA\_NGC1097 & $2932.6$ & $2932.6$ \\
             \hline
            1566 & T02\_085T01\_9000002296 & F154W & $2835.2$&$1.4$& &NGA\_NGC1566 &$3227$ & $3227$ \\
             \hline
            1672 &  A11\_003T10\_9000004838  & F148W &$2453.9$&$1.3$ & &NGA\_NGC1672 & $2690.1$ & $3548.1$  \\
             \hline
            3486&G08\_083T02\_9000001840  & F148W &$1728.4$& $1.7$ & &GI4\_095046\_NGC3486 & $2439.65$& $4023.65$\\
            \hline
            4051 &G05\_246T04\_9000000882  &F154W & $12552.5$&$1.5$& &AIS\_102 & $109$ & $109$ \\
             \hline
            5194 & T05\_023T01\_9000005058 &F148W & $11157.7$ &$1.3$& & GI3\_050006\_NGC5194 & $10787.45$ & $10787.45$ \\
            \hline
             \multicolumn{8}{l}{Non-AGN galaxies}\\
             \hline
            1300 &A07\_027T05\_9000003506 & F148W & $2179.4$&$1.6$& & AIS\_282 &$270$ & $270$\\
             \hline
            2903& G08\_031T03\_9000001972& F148W& $2995.8$&$1.3$& & NGA\_NGC2903& $1910.2$&$1909.2$\\
             \hline
            3351 & A09\_043T02\_9000003726 & F148W &$6574.9$& $1.2$& &NGA\_NGC3351 & $1692.2$ & $1692.2$\\
            & T03\_034T01\_9000002500  & & & & &\\
             \hline
            4321 &A08\_003T05\_9000003426 & F154W & $5368.7$ &$1.3$& &NGA\_NGC4321 &$1754.1$ & $2932.2$\\
             \hline

	\end{tabular}
       \vspace{0.1ex}

     {\raggedright \textsc{Notes.--}$^{\text{a}}$ Full Width at Half Maximum (FWHM) of the point sources in the FOV from UVIT observations.\par}
\end{table*}

\subsection{Data}
\subsubsection{{\sl AstroSat}-UVIT}

UVIT onboard {\sl AstroSat} consists of two Ritchey-Chretien (RC) telescopes. One telescope operates in the Far-UV (FUV; 130-180 nm), while the other covers the Near-UV (NUV; 200-300 nm) as well as the optical (VIS) band \citep[e.g.,][]{Kumar2012, Tandon2017, Tandon2020}. The UVIT provides a $14^{\prime}$ radius Field of View (FOV) alongside an angular resolution of $\sim 1.2$-$1.5^{\prime\prime}$. Observations presented here are carried out in photon-counting mode, with a readout rate of $\sim 29$ frames per second.

Level-1 data for {\sl AstroSat}-UVIT are obtained from the Indian Space Science Data Center (ISSDC)\footnote{https://astrobrowse.issdc.gov.in/astro\_archive/archive/Home.jsp}. We used the archival observations in the FUV BaF2 (F154W; $\lambda_{mean} = 1541\text{\AA}$) or FUV CaF2 (F148W; $\lambda_{mean} = 1481\text{\AA}$) filters of UVIT \citep{Tandon2020}. Table~\ref{tab:obs} provides the details of the observations used in this work. Level-1 data are processed through the CCDLAB \citep{Postma2021} pipeline to obtain Level-2 images and exposure maps. In the case of NGC 3351, two observations are present with the F148W filter. We combined these observations using \texttt{CCDLAB}. The final images of these ten galaxies have a plate scale of $\sim 0.416^{\prime\prime}$ and an angular resolution of 1.2-1.7$^{\prime\prime}$. The UVIT images of individual galaxies are provided in Figure~\ref{fig:UVIT_obs}.

\subsection{{\sl Galaxy evolution explorer (GALEX)}}
{\sl GALEX} has a modified RC telescope with a diameter of 50 cm observing in the FUV (1350-1750$\text{\AA}$; $\lambda_{eff} = 1538.6\text{\AA}$) and NUV (1750-2750$\text{\AA}$; $\lambda_{eff} = 2315.7\text{\AA}$) bands \citep{Martin2005, Morrissey2007}. The telescope has a FOV of 1.2$^\circ$ diameter with an angular resolution of 4.5-5$^{\prime\prime}$. 
Level-2 science-ready archival photometry observations from {\sl GALEX} in the FUV ($\lambda_{eff} = 1538.6\text{\AA}$) and NUV ($\lambda_{eff} = 2315.7\text{\AA}$) bands  are obtained from the MAST portal\footnote{https://galex.stsci.edu/gr6/}. The calibrated intensity and effective exposure map images with an angular resolution of 4.5-5$^{\prime\prime}$ are used for the analysis \citep{Morrissey2007}. We used the observation tiles having both FUV and NUV observations for a given galaxy. Details of the observation used are provided in Table~\ref{tab:obs}.

\section{Analysis and Results}\label{sec:Analysis}
\subsection{Astrometry}
Astrometry solutions for the UVIT observations are first attempted with \texttt{CCDLAB} \citep{Postma2020,Postma2021} using the Gaia catalogue \citep[][]{ 
Gaia2018}. If \texttt{CCDLAB} is unsuccessful, then we use \texttt{SCAMP} \citep[version 2.10.0;][]{Bertin2006} with the {\sl GALEX} catalogue (provided in the GR6/7 data release) to obtain the astrometry solution. For two galaxies, NGC 4051 and NGC 1300, \texttt{CCDLAB} and \texttt{SCAMP} failed to provide correct astrometry solutions. For these galaxies, we visually identified a few sources in the FOV against GALEX observations, and an astrometry solution is obtained using \texttt{ccmap} and \texttt{cctran} tasks in the Image Reduction and Analysis Facility (\texttt{IRAF})\footnote{https://iraf.net/}.

\subsection{UV Analysis and SFR calculation }\label{UV_knot}
We used a publicly available software package, \texttt{SExtractor} \citep{Bertin1996}, to identify the SF knots from UVIT images. The background and threshold are calculated using the method described in \citet{Ananthamoorthy2024}, which accounts for the Poisson distribution of UVIT astrophysical background. Source detection is carried out with a deblending parameter value (\texttt{DB$\_$MINCOUNT}) of $5 \times 10^{-7}$ as the region is crowded \citep[see, ][for details]{Bordoloi2024}.

We limited the analysis to the knots detected up to the R$_{25}$ of these galaxies. Flux measurements are carried out on background-subtracted images where the background is calculated in source-free regions outside the extent of the galaxy. We used FLUX\_AUTO from SExtrcator to compute the source flux, which contains $\sim$ 94\% of the total source flux \citep{Piridi2024, Bertin1996}. We converted the flux in counts per second (CPS) to erg cm$^{-2}$ s$^{-1}$ $\text{\AA}^{-1}$ using the unit conversion factor provided in \citet{Tandon2017}. Knots with high detection significance ($\gtrsim 5\sigma$) are only considered for further analysis to avoid any possible spurious detection due to lower \texttt{DB$\_$MINCOUNT}. To avoid contamination from the foreground stars, sources with NUV to FUV flux ratio (obtained from GALEX image as described later in this section) greater than 10 are removed \citep{Bigiel2008}.

The measured fluxes for the individual knots are corrected for the foreground Galactic extinction \citep{Cardeli1989} using the visual extinction values of \citet{Schlafly2011}. Further, they are corrected for internal attenuation using UV colour from {\sl GALEX} observation. To derive the UV colour, {\sl GALEX} FUV images are brought into NUV resolution by convolving the FUV images with the Gaussian convolution filter of $\sigma \sim 0.901$ pixels of {\sl GALEX}. The background, calculated outside the region of the galaxy in the FOV, is subtracted in both FUV and NUV images. The UV flux is measured in the box of 5 $\times$ 5 pixels ($\sim$ FWHM of {\sl GALEX}) centred on each knot detected from the UVIT observation. The box size is increased by another pixel in case of a negative value of CPS in either FUV or NUV images. The CPS is converted to the magnitude using the zero-point magnitudes provided in \citet{Morrissey2007}. 

\begin{table*}
	\centering
	\caption{Details of the global SF properties of the galaxies. \label{tab:sfr_prop}}
	
	\begin{tabular}{lcccccccc} 
		\hline
Name    & Knots     & Median            & MAD$^{\text{a}}$             &Median         & MAD$^{\text{a}}$          & Median            &MAD$^{\text{a}}$          & Total  \\
        & detected   & $\Sigma_{\text{SFR}}$              & $\Sigma_{\text{SFR}}$              & Area          & Area           & A$_{\text{FUV}}$              &A$_{\text{FUV}}$                   & SFR  \\
(NGC)   &          & (M$_\odot$yr$^{-1}$kpc$^{-2}$)      &  (M$_\odot$yr$^{-1}$kpc$^{-2}$) &    (kpc$^{2}$)         &   (kpc$^{2}$)  &        (mag.)           &  (mag.)                    &  (M$_\odot$yr$^{-1}$)     \\
\hline
\multicolumn{9}{l}{AGN-host galaxies} \\
\hline

1097   &$1463$	   &$0.080 \pm 0.005$	&$0.04$     &$0.0277$      &$0.022$       &	$1.289\pm0.007$    & $0.26$	     &$11.59\pm0.35$\\
1566	&$824$	   &$0.45 \pm 0.01$	    &$0.26$     &$0.0036$	   &$0.002$	      &$1.065\pm0.006$ 	   & $0.27$	     &$4.36\pm0.15$\\
1672	&$613$	   &$0.51\pm0.02$	    &$0.23$	    &$0.0049$	   & $0.0028$     &$1.390\pm0.008$	   &$0.20$	     &$6.15\pm0.23 $\\
3486	&$554$	   &$0.57\pm0.01$		&$0.29$	    &$0.0051$	   &$0.0025$	  &$1.210\pm 0.007$	   & $0.17$	     &$3.95\pm0.04$\\
4051	&$561$	   &$0.20\pm0.05$	    &$0.11$	    &$0.0117$	   &$0.0078$	  &$1.45\pm	0.03$	   &$0.40$	     &$4.2\pm0.3$\\
5194	&$1689$    &$0.51\pm0.02$		&$0.27$	    &$0.0041$	   &$0.0031$	  &$1.705\pm0.005$	   &$0.28$       &$10.04\pm0.12$\\
\hline
\multicolumn{9}{l}{Non-AGN Galaxies}\\
\hline
1300	&$327$	  &$0.35\pm0.02$		&$0.17$	    &$0.0078$	   &$0.0035$	   &$1.32\pm0.03$	   & $0.31$	 &$1.97\pm0.06$\\
2903	&$805$	  &$0.688\pm0.02$		&$0.49$	    &$0.0027$	   &$0.0017$	   &$1.541\pm0.008$	   &$0.41$	 &$4.78\pm0.09$\\
3351	&$634$	  &$0.38\pm 0.02$		&$0.19$		&$0.0034$	   &$0.0023$	   &$1.69\pm0.01$      &$0.29$	     &$2.61\pm0.17$\\
4321	&$616$	  &$0.33\pm 0.01$       &$0.15$		&$0.0123$	   &$0.0087$	   &$1.646\pm0.009$	   &$0.21$     &$7.23\pm0.23$\\
\hline

	\end{tabular}

    \vspace{0.1ex}

    {\raggedright \textsc{Notes.--} $^{\text{a}}$ Median Absolute Deviation for the distribution.\\
    The values in the table are for the entire galaxy. The regions affected by the merger are not excluded for NGC 1097 and NGC 5194.  \par}

\end{table*}

The UV spectral slope ($\beta$) is calculated from the {\sl GALEX} magnitudes using the relation, $\beta = [(m_1 - m_2)/(-2.5\log\frac{\lambda_1}{\lambda_2})] -2$ \citep{Nordon2013}, where $m_1$ and $m_2$ are the magnitudes in the GALEX FUV and NUV bands, respectively, and $\lambda_1$ and $\lambda_2$ are the corresponding wavelengths. The $\beta$ is used to calculate the colour excess, E(B-V), using the relation \citep{Reddy2018}, $\beta = -2.616+4.684\text{E(B-V)}$. Further, the Calzetti relation \citep{Calzetti2000}, as given below (where $\lambda$ is expressed in $\mu$m), is used to obtain the relative attenuation ($k_{\lambda}$) at the FUV wavelength.

\begin{equation}
    k_{\lambda} = 2.659\left[-2.156+\frac{1.509}{\lambda}-\frac{0.198}{\lambda^2}+\frac{0.011}{\lambda^3}\right] + R^{\prime}_{V}\;,
\label{Att_law}
\end{equation}

We considered a $R^{\prime}_V =4.05\pm0.8$ \citep{Calzetti2000}. Attenuation at UVIT wavelength (A$_{\text{FUV}}$) is calculated using the relation \citep{Calzetti2000}, A$_{\lambda} = (0.44\pm0.03)\text{E(B-V)}k_{\lambda}$. A small fraction of knots ($0.27\%$) yielded negative attenuation. For these knots, a median attenuation value of the corresponding galaxy is considered. The attenuation-corrected UV luminosity ($\text{L}_{\text{FUV}}$) is used to derive the SFR ($\text{SFR}_{\text{FUV}}$) using the following relation \citep{Iglesias2006, Cortese2008}. 
\begin{equation}
\text{SFR}_{\text{FUV}}(\text{M}_\odot \text{yr}^{-1}) = \frac{\text{L}_{\text{FUV}} (\text{erg s}^{-1})}{3.83 \times 10^{33}} \times 10^{-9.51}\;
\end{equation}

\subsection{Effect of mergers}
The galaxies NGC 1097 and NGC 5194 are interacting strongly with the neighbouring galaxies NGC 1097A and NGC 5195, respectively. Simulations and observations have suggested that mergers can trigger SF in the host galaxy \citep[e.g.,][ and references therein]{Bournaud2011}. The merger-induced impact on SF could be maximum at the regions of ongoing merger due to morphological disturbances \citep[e.g.,][]{Brandl2009} and in the nuclear regions due to driving of gas to the central region \citep[e.g.,][]{Barnes1991}. Therefore, we calculated the $\Sigma_{\text{SFR}}$ and other SF properties away from the merger-affected regions for these galaxies. We considered the SF knots detected in the semi-elliptical region (cut along the galaxy minor axis) away from the companion galaxy, as shown in Panels (a) and (f) of Figure~\ref{fig:UVIT_obs}. 

\begin{figure}
\centering
	\includegraphics[width=7cm]{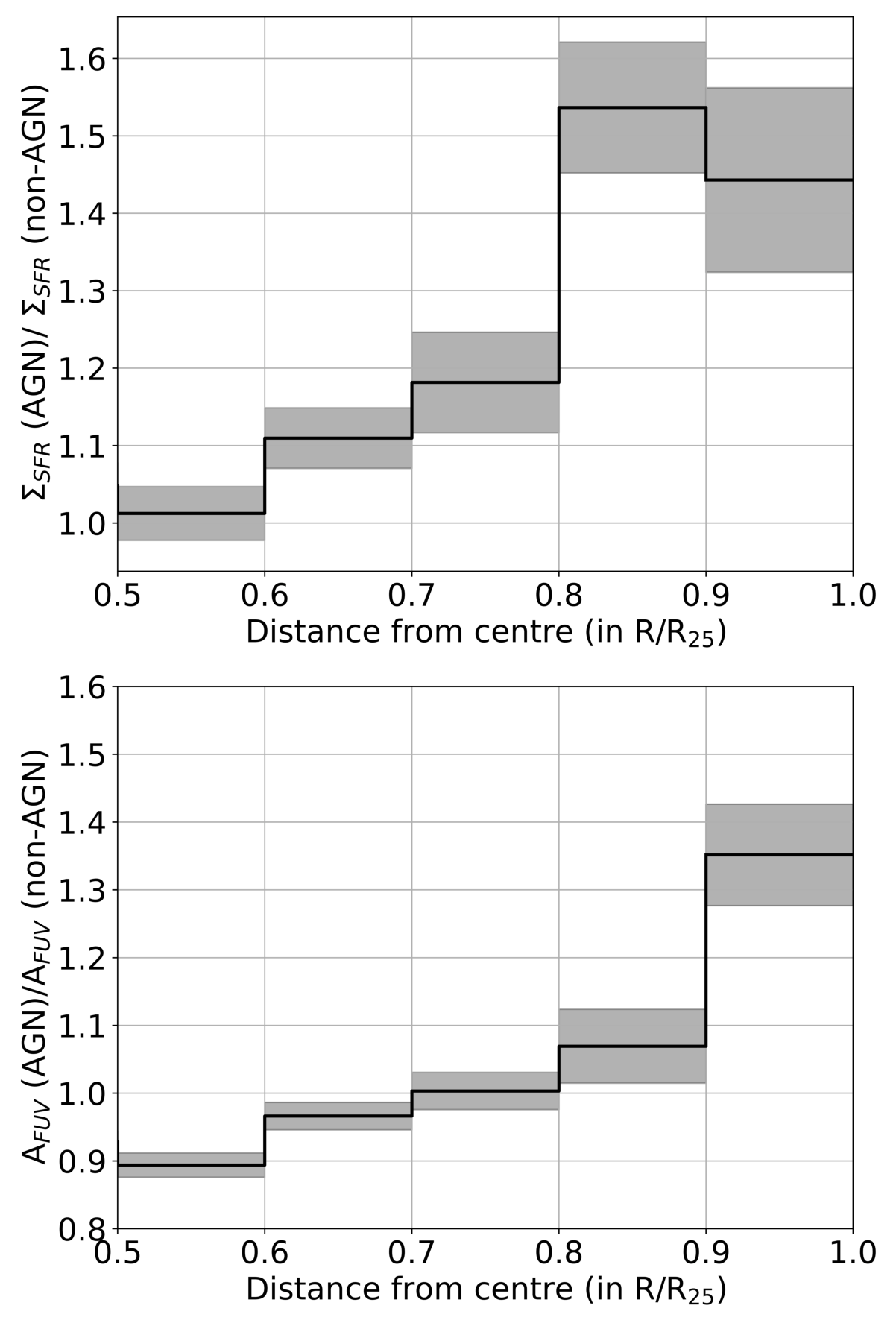}
   \caption{Upper Panel: Ratio of average $\Sigma_{\text{SFR}}$ of AGN-host and non-AGN galaxies as a function of distance from the centre. Lower Panel: Ratio of average A$_{\text{FUV}}$ of AGN-host and non-AGN galaxies as a function of distance from the centre.}
    \label{fig:sfrd_AFUV_ratio}
\end{figure}

\subsection{Size of the SF region and $\Sigma_{\text{SFR}}$}
The ellipse parameters, i.e., the semi-major axis and the semi-minor axis, obtained from \texttt{SExtractor}, are used to derive the area of the individual knots. The area is corrected for the angular resolution of the UVIT (FWHM for each observation is provided in Table-\ref{tab:obs}), considering that the knots have a spatial distribution of a 2D-elliptical Gaussian function \citep{Nandi2024}. For a fraction of faint UV knots, the size along the major and/or minor axis is less than the $\sigma$ of the UVIT Point Spread Function (PSF). It could be due to the poorer constraint on the fitted ellipse parameters in \texttt{SExtractor} for faint sources. For these sources, $\sigma$ of the UVIT PSF is considered as the knot size along that axis. The SFR is divided by the area for each knot to obtain the $\Sigma_{\text{SFR}}$ of the knot in the unit of M$_\odot$ yr$^{-1}$ kpc$^{-2}$ \citep[e.g.,][]{Kennicutt2012}. The number of knots detected, median area of the knots, $\Sigma_{\text{SFR}}$, and total SFR of individual galaxies are provided in Table~\ref{tab:sfr_prop}. The SFR values provided in Table~\ref{tab:sfr_prop} can differ from those available in the existing studies. One such example is that \citet{Song2021} quoted the UV SFR value as $0.69\pm0.03$ M$_\odot$ yr$^{-1}$ for NGC 1097 (in Table~5) in contrast to $11.59\pm0.35$ M$_\odot$ yr$^{-1}$ estimated in this work. We notice that intrinsic attenuation correction is the primary factor that was responsible for this difference in the SFR.

\subsection{$\Sigma_{\text{SFR}}$ and A$_{\text{FUV}}$ in the outer regions of the galaxy:}

The UV SFR of knots is divided by the total area of knots in a radial bin to obtain $\Sigma_{\text{SFR}}$ as a function of the distance (relative to $R_{25}$) from the galaxy centre. It is to be noted that though the derived SFR depends on the distance to the galaxy, the $\Sigma_{\text{SFR}}$ is independent of the distance. In this paper, all references to the radial bin are relative to $R_{25}$, and also knots corresponding to each radial bin are extracted in an elliptical annular region accounting for the inclination angle of the galaxy. The $\Sigma_{\text{SFR}}$ at different radial bins is averaged for different galaxies (separately for AGN-host and non-AGN galaxies) to obtain the average $\Sigma_{\text{SFR}}$ as a function of the distance from the centre. The ratio of the average $\Sigma_{\text{SFR}}$ for AGN-host and non-AGN galaxies is provided in the Upper Panel of Figure~\ref{fig:sfrd_AFUV_ratio}, which clearly shows the enhancement in $\Sigma_{\text{SFR}}$ for UV SF knots at radius $>0.6 R_{25}$ of AGN-host galaxies. Therefore, in comparison to the non-AGN galaxies, a significant enhancement of $\Sigma_{\text{SFR}}$ on the outskirts of the AGN-host galaxies suggests a possible positive AGN feedback.

We also derive radial profiles of A$_{\text{FUV}}$ for individual galaxies using the median A$_{\text{FUV}}$ of the knots in each radial bin. Next, we estimate the A$_{\text{FUV}}$ in each radial bin averaged over (a) all AGN-host and (b) all non-AGN galaxies. The ratio of A$_{\text{FUV}}$ averaged over all AGN-host galaxies to that of all non-AGN galaxies is provided in the Lower Panel of Figure~\ref{fig:sfrd_AFUV_ratio}. It suggests that AGN-host galaxies have higher dust in the outer region compared to non-AGN galaxies, most likely due to AGN feedback.

We further examine the rate of decrease in $\Sigma_{\text{SFR}}$ and A$_{\text{FUV}}$ in the outer regions ($>0.5$R$_{25}$) of each galaxy by fitting a linear function of the form $X(r) = ar+b$, where X is $\log_{10}(\Sigma_{\text{SFR}})$ and A$_{\text{FUV}}$ respectively.  

\begin{figure*}
\centering
	
	\includegraphics[width=10cm]{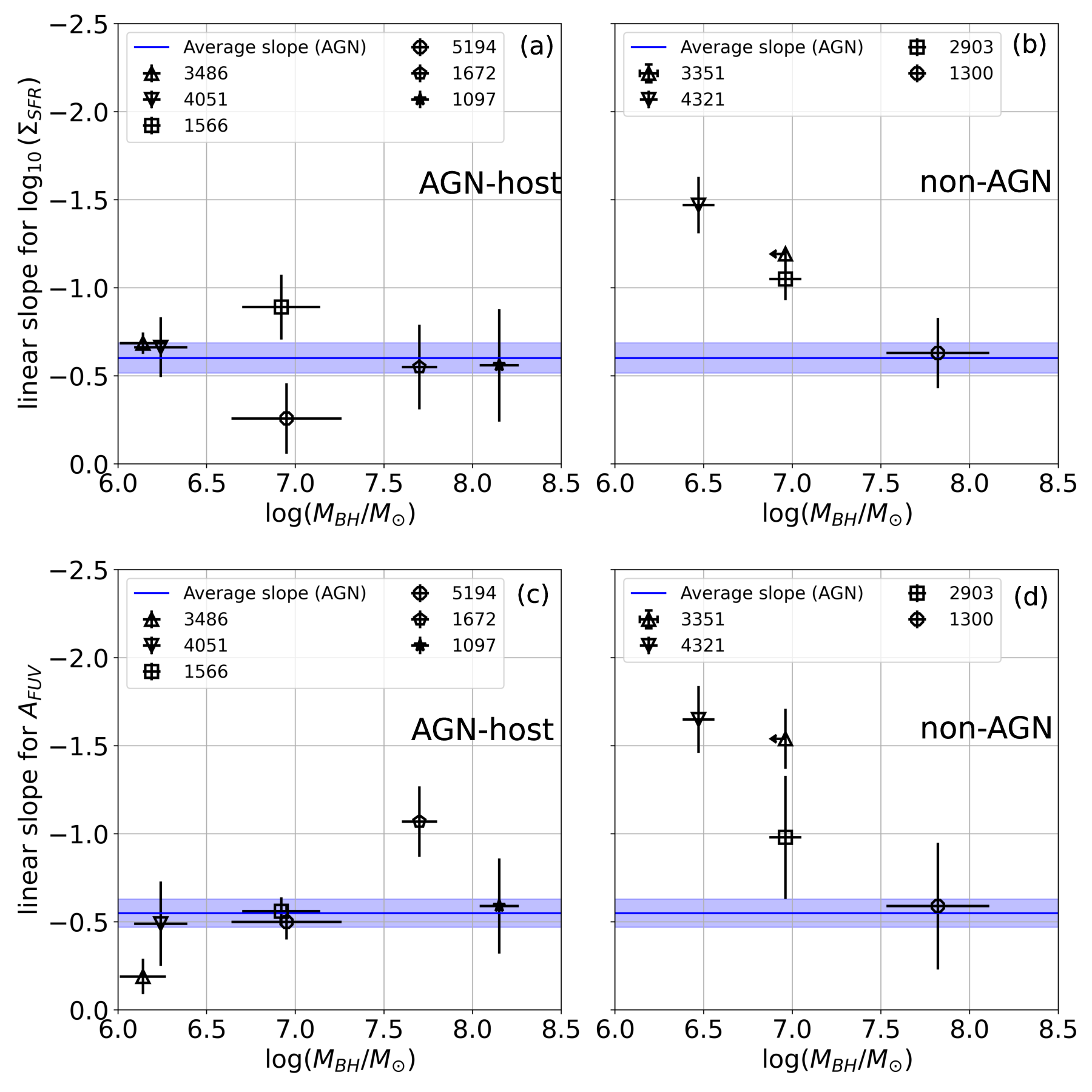}
  \caption{Panel (a): Slopes of straight lines fitted to the radial profile of $\log_{10}(\Sigma_{\text{SFR}})$ for individual AGN galaxies in the galaxy outskirts as a function of M$_{BH}$. The mean slope for AGN galaxies and the corresponding 1-$\sigma$ error (shaded region) are provided for reference. Panel (b): Same as Panel (a) but for non-AGN galaxies. The mean slope for AGN galaxies is provided for reference. Panel (c): Slopes of straight lines fitted to the radial profile of A$_{\text{FUV}}$ for individual galaxies in the galaxy outskirts as a function of M$_{BH}$. Panel (d): Same as Panel (c) but for non-AGN galaxies.}
    \label{fig:slope_individual_gal}
\end{figure*}

It is to be noted that the radial profiles for all the galaxies are not strictly linear. However, this parametric approach provides the overall radial trend in $\Sigma_{\text{SFR}}$ and A$_{\text{FUV}}$ of a galaxy. In the case of $\Sigma_{\text{SFR}}$, the slopes of the individual AGN and non-AGN galaxies are provided in Figure~\ref{fig:slope_individual_gal}-(a) and Figure~\ref{fig:slope_individual_gal}-(b), respectively. The average slope for AGN-host galaxies is $0.60\pm0.09$, as shown in the shaded region on Figure~\ref{fig:slope_individual_gal}-(a) and Figure~\ref{fig:slope_individual_gal}-(b). We notice that the non-AGN galaxies have relatively steeper individual slopes than the average slope observed in AGN-host galaxies. We also noticed that this trend exists without correcting the $\Sigma_{\text{SFR}}$ for intrinsic extinction correction. This effect is found to be prominent for galaxies hosting relatively lower mass central black holes. This finding suggests that $\Sigma_{\text{SFR}}$ of knots in the outskirts of AGN-host galaxies, particularly for the lower black hole mass systems, fall off less rapidly than in non-AGN galaxies. Further, for non-AGN galaxies, we notice a possible flattening of the slopes with increasing central black hole mass. However, given the small sample size (4 galaxies), the statistical significance of this trend is not strong.

The slopes of the radial profile of A$_{\text{FUV}}$ for individual AGN and non-AGN galaxies are shown in Figure~\ref{fig:slope_individual_gal}-(c) and Figure~\ref{fig:slope_individual_gal}-(d), respectively. The shaded region indicates the average slope ($0.55\pm0.08$) for the AGN-host galaxies. Similar to $\Sigma_{\text{SFR}}$, A$_{\text{FUV}}$ slopes in the AGN-host galaxies are flatter than those in the non-AGN galaxies. Here, we also noticed that the effect is primarily seen in the systems with relatively lower-mass central black holes. 

We propose that due to the positive AGN feedback, more gas and dust were moved to the outskirts of the AGN-host galaxies than the non-AGN ones. Further, in our sample, the AGN feedback is prominent in relatively lower-mass black hole systems.

\subsection{Kolmogorov–Smirnov and Anderson-Darling test:}
We carried out the commonly used 2-sample Kolmogorov-Smirnov (KS) \citep[e.g.,][]{Rizzo2007} to check if differences in slope values of $\Sigma_{\text{SFR}}$ and A$_{\text{FUV}}$ between AGN and non-AGN are statistically significant. We used the \texttt{ks$\_$2samp} routine in \texttt{scipy.stats} package in \texttt{Python} to carryout the analysis. Under the null hypothesis that the distribution for AGN and non-AGN is equal, we obtained a p-value of 0.048 for $\Sigma_{\text{SFR}}$ slopes. The test favours the alternative hypothesis that ``AGN have flatter slopes than non-AGN'' at a confidence level $\sim 95\%$. In addition, we also used a 2-sample Anderson-Darling (AD) test \citep{Anderson1954, Pettitt1976} to compare the distributions. We used the \texttt{anderson$\_$ksamp} routine in \texttt{scipy.stats} package for this analysis. The AD statistic also yielded a similar confidence level ($\sim 96\%$), which marginally favours the alternative hypothesis. For A$_{\text{FUV}}$, the p-value is between $0.023-0.033$ for both tests, which also marginally favours the alternative hypothesis.

\section{Discussion and Conclusions}
 
\subsection{Effect of bar and mergers}\label{sec:bar_effect}
All galaxies in our sample are barred, and bar strength can have an impact on the SF of the host galaxy \citep{Robinchaud2017}. Therefore, we obtained the bar strength parameter ($Q_b$) for the galaxies in our sample from the earlier works, which are provided in Table~\ref{tab:bar_asym}. Though our sample size is limited, we did not observe any significant dependence of the slope of $\Sigma_{\text{SFR}}$ with $Q_b$, suggesting that the bar effect on SF of the outer regions might be minimal in our sample.

\begin{table}

	\centering
	\caption{Bar strength and Asymmetry parameter for the galaxies. \label{tab:bar_asym}}
	
	\begin{tabular}{lccc} 
		\hline
Name    & Bar Strength     & Asymmetry            & Ref.            \\
(NGC)   & $Q_{b}$           & (A)              & $Q_{b}$            \\              
\hline
\multicolumn{4}{l}{AGN-host galaxies} \\
\hline

1097    &$0.26$	       &$0.65$	&$(a)$     \\
1566	&$0.153$	   &$0.68$	    &$(b)$     \\
1672	&$0.37$	       &$0.88$	    &$(a)$	    \\
3486	&$0.155$	   &$0.57$		&$(c)$	    \\
4051	&$0.097$	   &$0.88$	    &$(d)$	    \\
5194	&$0.165$       &$0.74$		&$(c)$	    \\
\hline
\multicolumn{4}{l}{Non-AGN Galaxies}\\
\hline
1300	&$0.57$	       &$0.84$		&$(a)$	    \\
2903	&$0.26$	       &$0.55$		&$(c)$	    \\
3351	&$0.24$	       &$0.15$		&$(a)$		\\
4321	&$0.21$	       &$0.88$       &$(c)$		\\
\hline

	\end{tabular}\\
{\raggedright \textsc{Notes.--} References: a: \citet{Cisternas2013}; b: \citet{Buta2009}; c: \citet{Laurikainen2002}; d: \citet{Buta2005}. \par}

\end{table}
Mergers can also significantly impact the SF of the host galaxy. Although we excluded the region with close interaction with the nearby galaxy for NGC 1097 and NGC 5194, the merger can still influence the global SF of the galaxy \citep[e.g.,][]{Moreno2021}. We used the non-parametric asymmetry parameter \citep[A;][]{Rodriguez2019, Conselice2000} for the galaxies in our sample to look for a possible merger effect. Galaxies undergoing a merger are expected to have a larger asymmetry due to disturbed morphology \citep[e.g.,][]{Conselice2000}. We used \texttt{STATMORPH} routine \citep{Rodriguez2019} in \texttt{Python} for deriving the asymmetry parameter using UVIT observations. As galaxies are well resolved in UV with many substructures, we noticed that the extracted asymmetry parameter could be driven to local minima \citep{Conselice2000}, yielding a higher value. To avoid such an overestimation, substantial smoothing of the image was essential. We smoothed the UVIT images with a Gaussian convolution filter of $\sim 3$ FWHM (9 pixels). The parameter values obtained are provided in Table~\ref{tab:bar_asym}. The values obtained are comparable with the asymmetry parameters derived in earlier studies using UV observation \citep{Holwerds2011}. We did not notice any significant difference in the asymmetry parameter value for NGC 1097 and NGC 5194 (interacting systems) compared to the other galaxies. We also did not observe any significant dependence on this parameter with a slope of $\Sigma_{\text{SFR}}$. Therefore, mergers may not be driving/biasing the difference in the radial profiles of AGN and non-AGN galaxies in our sample.

\subsection{Possible mechanism for enhanced $\Sigma_{\text{SFR}}$ in the outskirts of AGN-host galaxies}
The enhanced $\Sigma_{\text{SFR}}$ for UV knots in the outskirts of AGN-host galaxies compared to non-AGN galaxies could indicate AGN-triggered SF in the outskirts of the galaxies. The observed relatively flatter slope of $\Sigma_{\text{SFR}}$ in individual AGN-host galaxies further supports this claim. Our analysis also shows that the outer regions of the AGN-host galaxies have higher A$_{\text{FUV}}$ compared to the non-AGN galaxies, indicating that the feedback might be coupled with dust. Furthermore, the observed relatively flatter slope of A$_{\text{FUV}}$ in individual AGN-host galaxies supports this proposition. The statistical tests also marginally favour the different radial profiles for AGN and non-AGN galaxies.

The enhanced specific SFR on the outskirts of AGN-host galaxies was observed by \citet{Sanchez2018} and \citet{Acharya2024}. \citet{Sanchez2018} also noticed the flatter gas surface density in the AGN-host galaxies. However, the gas surface density was derived from visual extinction, which is also a good tracer of dust. Therefore, this finding also indicates a higher presence of dust in the outskirts of galaxies, which is consistent with our observation. 

As the AGN-host galaxies in our sample are weak-jetted, the prominent AGN feedback mechanism could be via the radiative mode. However, the L$_{bol, AGN}$ of AGN (provided in Table~\ref{tab:gal_prop}) in our sample is lower than the typical critical luminosity ($\sim 10^{46}$erg s$^{-1}$) for launching outflow \citep{Costa2018, Ishibashi2012}. This low luminosity suggests the current activity state may not be able to drive strong outflows in the galaxy. Therefore, if the observed effect is due to AGN feedback, then it could be by ``fossil outflows'' as suggested by \citet{Ishibashi2018}, where past powerful AGN activity/episodic events could have driven outflows. Overall results of this study are consistent with ``the SFR enhancement correlated with fading of AGN'' scenario as suggested in simulations \citep{Zubovas2017}. 

It should be noted that the temporal luminosity evolution of AGN is still uncertain. With simplest models like exponential decay \citep{Ishibashi2018}, a decay time scale of a few Myr is essential to explain the enhanced $\Sigma_{\text{SFR}}$ with fading of AGN \citep[assuming stellar populations have an age of a few $10$ Myr;][]{Sanchez2011}. This value is comparable to typically used/derived AGN activity time scales \citep[1 Myr;][]{Hickox2014, Ishibashi2018, Harrison2024}. However, future multiband observations at a similar spatial scale and associated stellar populations modelling to estimate their age are essential to understand the time scales of SF to connect with the AGN activity time scale.

The possible mode through which AGN feedback can influence the SF is via radiation-pressure-driven AGN feedback via dust or radiative wind \citep[e.g.,][]{King2015, Ishibashi2012}. The simulations by \citet{Costa2018} suggest that the radiation-pressure-driven AGN feedback via dust can drive outflows. According to \citet{Ishibashi2012} and \citet{Ishibashi2013}, the radiation pressure-driven feedback can drive the shells of dust and gas towards the outer region of a galaxy. SF can occur in the outer regions of the galaxy, obscured by dust, resulting in enhanced SF and dust in the outer regions. Even wind-mode feedback can enhance the SF on the outskirts of the galaxy. It is suggested that wind-mode feedback can remove gas from the galaxy bulge, but may trigger SF in the galaxy disk \citep[e.g.,][and references therein]{King2015}.

In this work, we cannot rule out the possibility of both wind and radiation-pressure-driven via dust modes of AGN feedback. However, as radiation-pressure-driven AGN feedback via dust mode preferentially couples with the dust, the study of dust-to-gas density ratio, extending up to R$_{25}$, can provide crucial insight into the mode of AGN feedback. However, it is to be noted that, currently, resolved molecular gas observations extending up to R$_{25}$ are very limited for nearby galaxies.

\subsection{Possible contribution from AGN ionisation}
Studies on nearby Seyfert galaxies suggest that the ionisation region from AGN can extend up to only $\sim 1-2$ kpc \citep[e.g.,][]{Schmitt2003}. It is to be noted that our study focuses on the regions more than $5$ kpc away from the central AGN. As per our knowledge, only one of the galaxies in our sample, i.e.,  NGC 4051, hosts a kpc-scale AGN ionisation region \citep{Meena2021}. This ionisation region is $\sim 1$ kpc in size around the central AGN, much smaller than the region probed ($5-10$ kpc) in this work. Further, the extent of the ionisation region in AGN was also observed to be dependent on the AGN luminosity \citep[e.g.,][]{Chen2019, Husemann2022}. We used the relation provided in equation~4 of \citet{Husemann2022} to estimate the minimum AGN luminosity required for producing the ionisation region of size corresponding to the smallest distance scale ($\sim5$ kpc) probed in this study. The estimated luminosity was $\sim 10^{44}$ ergs s$^{-1}$, which is one order higher than the maximum bolometric AGN luminosity ($\sim 6 \times 10^{42}$ ergs s$^{-1}$) of the galaxies in our sample. Therefore, we conclude that the contribution to UV emission due to AGN ionisation is expected to be negligible.

\subsection{Summary:}
We carried out a detailed study of SF properties in the outer regions of ten (six AGN + four non-AGN) galaxies using {\sl AstroSat}-UVIT and {\sl GALEX} observations. High-resolution {\sl AstroSat}-UVIT observations provided the best platform as of today for the comprehensive comparison of SF properties in nearby AGN and non-AGN galaxies extending up to R$_{25}$ with a spatial resolution of 50-100 pc. Though the number of galaxies is limited, considering the non availability of improved high-resolution UV images for the next many years, we conclude the following: The $\Sigma_{\text{SFR}}$ and A$_{\text{FUV}}$ in the outer regions of the AGN-host galaxies are enhanced and falls off less rapidly compared to the non-AGN galaxies, suggesting a positive AGN feedback in the outer region of the AGN-host galaxies. We speculate that the radiation pressure on dust and/or wind-driven AGN feedback could be playing a role in enhanced $\Sigma_{\text{SFR}}$ for AGN-host galaxies. Future deep multiband observations with the $\sim1^{\prime\prime}$ resolution or better of a larger galaxy sample are required to identify the exact mode of this observed AGN feedback.

\begin{acknowledgement}

We thank the anonymous reviewer and the associate editor for their constructive comments and suggestions that improved the manuscript. BA acknowledges the financial support by DST, Government of India, under the DST-INSPIRE Fellowship (Application Reference Number: DST/INSPIRE/03/2018/000689; INSPIRE Code: IF190146) program. DB thanks the Department of Space, Govt of India, for the financial support under the ISRO-RESPOND project (project no: RES-URSC-2023-031, sanction order number: No.DS-2B-13012(2)/10/2025-Sec.2). The research is based to a significant extent on the results
obtained from the AstroSat mission of the Indian Space Research
Organisation (ISRO), archived at the Indian Space Science Data Center
(ISSDC). The Payload Operations Centre at IIA processed the UVIT data used here. The UVIT is
built in collaboration between IIA, IUCAA, TIFR, ISRO and CSA. This research is based on observations made with the {\sl Galaxy Evolution Explorer}, obtained from the MAST data archive at the Space Telescope Science Institute, which is operated by the Association of Universities for Research in Astronomy, Inc., under NASA contract NAS 5–26555. We acknowledge the usage of the HyperLeda database (http://leda.univ-lyon1.fr). This research has made use of the SIMBAD database, CDS, Strasbourg Astronomical Observatory, France; VizieR catalogue access tool, CDS, Strasbourg Astronomical Observatory, France (DOI: 10.26093/cds/vizier); NASA/IPAC Extragalactic Database (NED), which is funded by the National Aeronautics and Space Administration and operated by the California Institute of Technology. The work has made use of the following software packages: Source Extractor \citep{Bertin1996}, CCDLAB \citep{Postma2017, Postma2021}, SAOImageDS9 \citep{ds92003}, Matplotlib \citep{Matplotlib2007}, Astropy \citep{Astropy2013,Astropy2018},  photutils \citep{photutils}, IRAF \citep{Tody1993}, STATMORPH \citep{Rodriguez2019}. Manipal Centre for Natural Sciences, Centre of Excellence, Manipal Academy of Higher Education (MAHE) is acknowledged for facilities and support.
\end{acknowledgement}

\section*{Data Availability}
The UVIT Level-1 data used in this work are publicly available and can be obtained from \url{https://astrobrowse.issdc.gov.in/astro_archive/archive/Home.jsp}. The {\sl GALEX} data described here may be obtained from the MAST archive at
\url{https://dx.doi.org/10.17909/hhjg-px52}. The other analysed data presented in this study can be made available upon reasonable request to the corresponding authors.

\bibliography{Astro_ref}

\end{document}